\documentstyle[aps,prl,twocolumn,floats,epsf]{revtex}

\begin{document}

\wideabs{
\title{New insulating phases of two-dimensional electrons in high
Landau levels: observation of sharp thresholds to conduction}

\author{K.~B. Cooper$^1$, M.~P. Lilly$^1$, J.~P. Eisenstein$^1$, 
L.~N. Pfeiffer$^2$, and K.~W. West$^2$}

\address{$^1$California Institute of Technology, Pasadena CA 91125 \\
         $^2$Bell Laboratories, Lucent Technologies, Murray Hill, NJ 07974}

\maketitle

\begin{abstract}
The intriguing re-entrant integer quantized Hall states recently discovered in
high Landau levels of high-mobility 2D electron systems are found to exhibit
extremely non-linear transport.  At small currents these states reflect
insulating behavior of the electrons in the uppermost Landau level.  At larger
currents, however, a discontinuous and hysteretic transition to a conducting
state is observed.  These phenomena, found only in very narrow magnetic field
ranges, are suggestive of the depinning of a charge
density wave state, but other explanations can also be constructed.   
\end{abstract}

}

Two-dimensional electron systems (2DES) in strong magnetic fields have
proven to be a remarkably rich laboratory for many-body 
physics\cite{perspectives}. The
continuing improvements in the techniques for creating 2DES in
semiconductor heterostructures have been paralleled by a steady stream
of discoveries of novel electron correlation phenomena. While the
fractional quantum Hall effect (FQHE) in the lowest ($N=0$) Landau level
(LL) is the best known of these, there has been a recent realization
that interactions among electrons in the excited LLs can give
rise to whole new classes of many-body phenomena. For example, recent
transport measurements\cite{lilly_ani,du_ani} have revealed huge and unexpected
anisotropies of the resistivity of the 2DES when the third and higher
($N \ge 2$) LLs are half filled. These anisotropies are not seen in the lowest
two LLs and appear only at very low temperatures and in the highest
quality samples. The observations are in qualitative agreement with
earlier theoretical suggestions of unidirectional charge density wave
(CDW) ground states (``stripe phases'') in the half-filled $N \ge 2$ 
LLs\cite{fogler,moessner}.
More recent theoretical work\cite{fradkin,fertig,yang,phillips}, 
going beyond the Hartree-Fock
approximation, has lent support to the stripe phase picture, albeit with
possibly important modifications due to quantum fluctuations. 

The experiments in high Landau levels also reveal remarkable phenomena
away from half filling, in the flanks of the LLs. In this regime both
Lilly, {\it et al.}\cite{lilly_ani} and Du, {\it et al.}\cite{du_ani}
reported the resistivity to be essentially isotropic and to fall to zero
in narrow regions of magnetic field near $\frac{1}{4}$ and $\frac{3}{4}$
filling of the LLs. In these regions the Hall resistance is found to be
accurately quantized but, quite surprisingly, {\it at the value of the
adjacent integer quantum Hall plateaus}. These re-entrant integer quantum
Hall effect (RIQHE)
states, which have only been found in the $N \ge 2$ LLs, suggest the
existence of insulating phases of the electrons in the uppermost LL. In
this paper we report the observation of a discontinuous transition from
the insulating state to a conducting one when large electric fields are
applied. This transition is found to be hysteretic and extremely
temperature and magnetic field dependent. The results are suggestive
of the depinning of CDWs\cite{gruner}, but they are also reminiscent of
quantum Hall breakdown phenomena\cite{nachtwei}. 

The samples used in this investigation are modulation-doped GaAs/AlGaAs
heterostructures grown by molecular beam epitaxy. Data from two samples
(A and B) are discussed here. Sample A exhibits a 2D electron density of
$n_s=2.7 \times 10^{11}$cm$^{-2}$ and a low temperature mobility of 
$1.1 \times 10^7$cm$^2$/Vs. Sample
B has a density of $n_s=2.1 \times 10^{11}$cm$^{-2}$ and a mobility of 
$1.6 \times 10^7$cm$^2$/Vs.
These parameters are determined after brief illumination with red light
at low temperature. Each sample consists of a square cleaved from
its parent wafer along 
$\langle 1 1 0 \rangle$ and $\langle 1 \overline{1} 0 \rangle$ crystal directions. 
Sample A is
$5 \times 5$mm$^2$ while sample B is $2.5 \times 2.5$mm$^2$. Eight indium
ohmic contacts are placed at the corners and midpoints of the sides of
the square. Longitudinal resistance measurements are performed by
injecting and withdrawing current from two midpoint contacts on opposite
sides of the square sample while recording the voltage difference between two
corner contacts. 
Both direct and low frequency alternating
currents have been employed in these experiments.

\begin{figure}[t]
\begin{center}
\epsfxsize=3.3in
\epsfclipon
\epsffile[69 141 536 481]{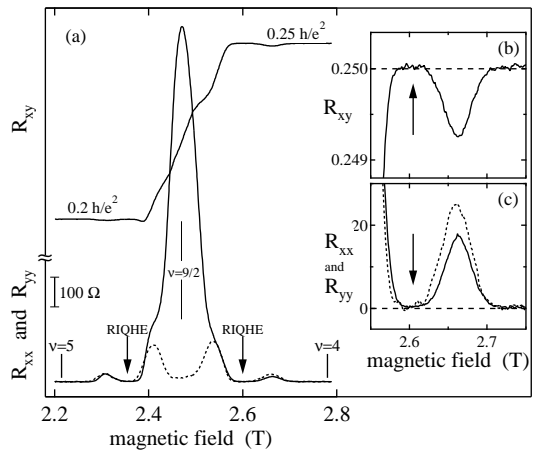}
\end{center}
\caption[figure 1]{(a) Longitudinal (solid line: $R_{xx}$, dotted line: $R_{yy}$) 
and Hall ($R_{xy}$) resistance 
of sample A in the $N=2$ Landau level at $T=50$mK.  
Arrows indicate the positions
of the re-entrant integer quantum Hall effect states.  Insets (b) and (c) magnify 
the RIQHE region.}
\end{figure}

Figure 1a shows transport data from sample A at $T=50$mK taken using an excitation
current of 10nA at 13Hz. The data cover magnetic
fields $B$ between LL filling fractions $\nu \equiv hn_s/eB=5$ and
$\nu=4$. In this range the Fermi level of the 2DES is in the lower spin
branch of the $N=2$ LL. The solid curves
are the longitudinal and Hall resistances, $R_{xx}$ and $R_{xy}$, 
observed for net
current flow along the $\langle 1 \overline{1} 0 \rangle$ direction. The
dotted curve is the longitudinal resistance $R_{yy}$ resulting from
net current flow along the $\langle 1 1 0 \rangle$ direction. The giant
anisotropy of the resistance near half filling of high Landau levels
reported earlier\cite{lilly_ani,du_ani} is clearly evident. 
This transport anisotropy, which is not seen in the 
$N=0$ or 1 LL, disappears above about 150mK. In the same field range 
no quantized plateaus appear in $R_{xy}$. The shoulders visible in Fig. 1a 
on either side of $\nu=9/2$ are not nascent plateaus.  They shift, and
ultimately disappear, as the temperature is lowered to $T=25$mK. 

The data in Fig. 1a also show that the resistance becomes approximately 
isotropic in the flanks of the Landau level. Furthermore, there are 
clearly defined regions of magnetic field (indicated by arrows) on each 
side of $\nu=9/2$ in which both 
$R_{xx}$ and $R_{yy}$ have dropped (in an approximately thermally
activated manner\cite{du_ani}) to vanishingly small values. These features, which
suggest the existence of fractional quantized Hall states, occur near
filling factors $\nu \approx 4 \frac{1}{4}$ and $4 \frac{3}{4}$. Measurements of
the Hall resistance $R_{xy}$ do indeed show quantization, but at the
value of the nearest {\it integer} quantized Hall plateau. Figures 1b and 1c
demonstrate this via magnified views of the resistances. The
data show that the RIQHE states are separated
from the main integer states by narrow regions of field in which
$R_{xx}$ (and $R_{yy}$) is non-zero and $R_{xy}$ is not
quantized\cite{lilly_ani,du_ani}. In common with the transport
anisotropies at half filling, these intriguing QHE features are seen in
several high LLs but they are conspicuously absent in the $N=1$ and 0
level. 

In the standard picture of the integer QHE, the finite width of Hall
plateaus and zero resistance states is attributed to the localization,
via disorder, of the quasiparticles in the system. Moving away from
integer filling increases the density and localization length of these
quasiparticles. When they eventually delocalize, the quantization of $R_{xy}$
is lost and $R_{xx}$ becomes non-zero. The data in Fig.1 show, quite
remarkably, that these same quasiparticles apparently localize again at
slightly higher densities where the RIQHE forms. This result cannot be 
readily understood in the standard single-particle 
localization picture of the integer QHE but instead suggests that these
new insulating phases in high LLs are critically dependent upon
electron-electron interactions. In the absence of disorder when the uppermost 
LL is just beginning to fill, interactions should lead to Wigner crystallization. 
With weak disorder this crystal would be pinned and therefore non-conducting.  
As the level is further filled, the Wigner crystal presumably melts. Interestingly,
recent theoretical works predict the existence, in high LLs, of additional ordered 
phases 
(``bubbles''\cite{fogler,moessner} and ``stripe crystals''\cite{fradkin}) between this 
Wigner crystal at the very edge of the LL and the unusual anisotropic 
phases at half filling. Conceivably, the insulating behavior that we observe in 
the RIQHE reflects the pinning of one of these new phases.

\begin{figure}
\begin{center}
\epsfxsize=3.3in
\epsfclipon
\epsffile[100 42 455 258]{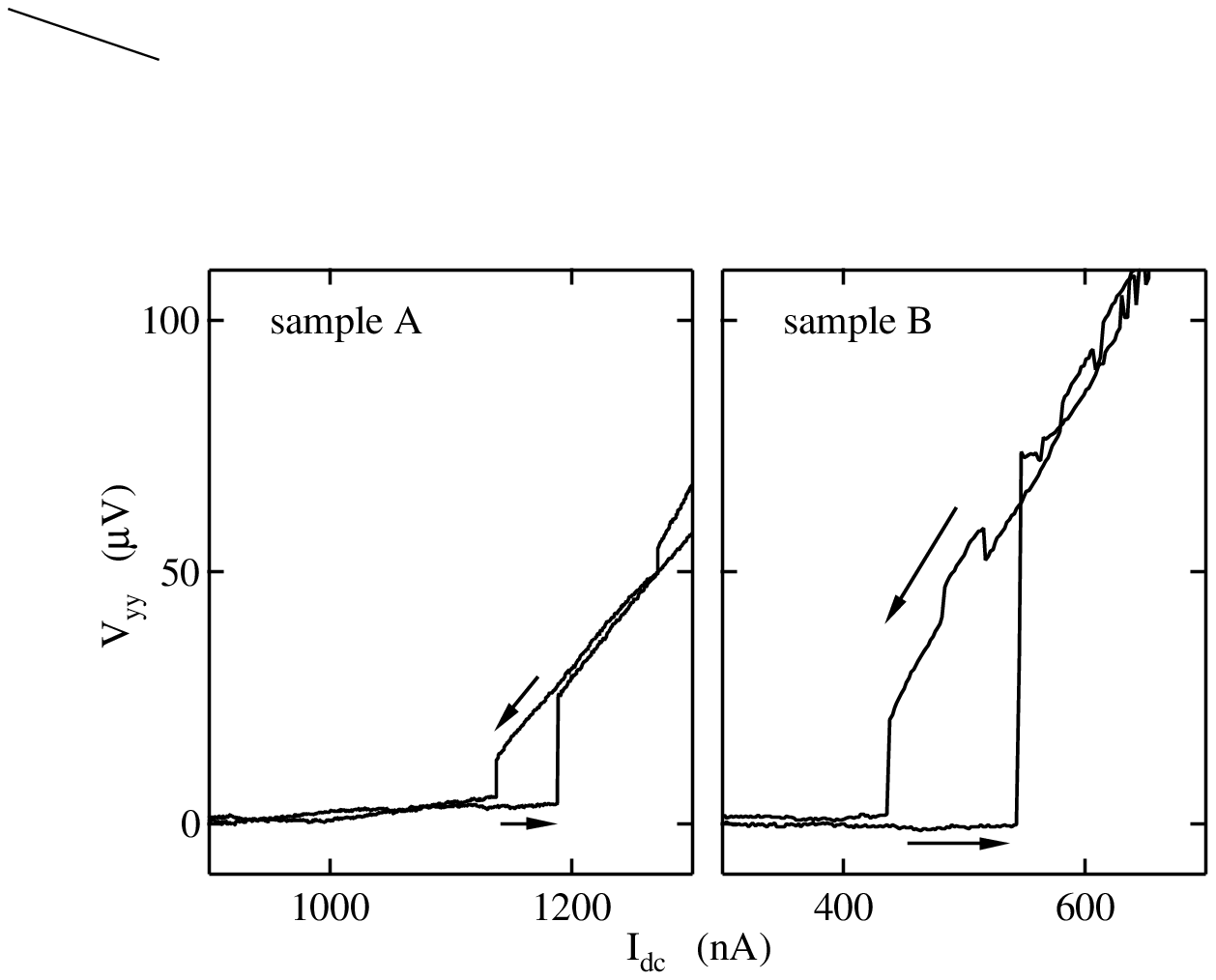}
\end{center}
\caption[figure 2]{Discontinuous current-voltage characteristics observed 
at $T = 25$ mK in the center of the RIQHE near $\nu \approx 4 \frac{1}{4}$.
Arrows denote direction of current sweep.}
\end{figure}

To examine this possibility, we have performed dc current-voltage (I-V)
studies of the RIQHE states. Figure 2 shows representative results from
both samples. The data shown come from the RIQHE region
around $\nu \approx 4 \frac{1}{4}$ at $T=25$mK. The dc current (along $\langle 1
1 0\rangle$) is slowly swept up from zero to as much as 1500nA and then
swept down again while the longitudinal voltage is recorded. 
In both samples a sharply defined threshold current is found where
the voltage jumps discontinuously. 
The transitions are hysteretic; the threshold current is almost 
always larger when the current is swept up than when it is swept down. 
Just above threshold the voltage rises approximately linearly with current, 
although additional small features can be seen. For sample A the threshold
is around 1200nA; for sample B it is around 500nA. The precise threshold 
current also depends sensitively upon the temperature and magnetic field within 
the RIQHE region.  No qualitative dependence of the thresholds 
upon the current flow direction or voltage probe configuration has been found, 
although quantitative variations are indeed observed.  

\begin{figure}
\begin{center}
\epsfxsize=3.3in
\epsfclipon
\epsffile[66 74 272 300]{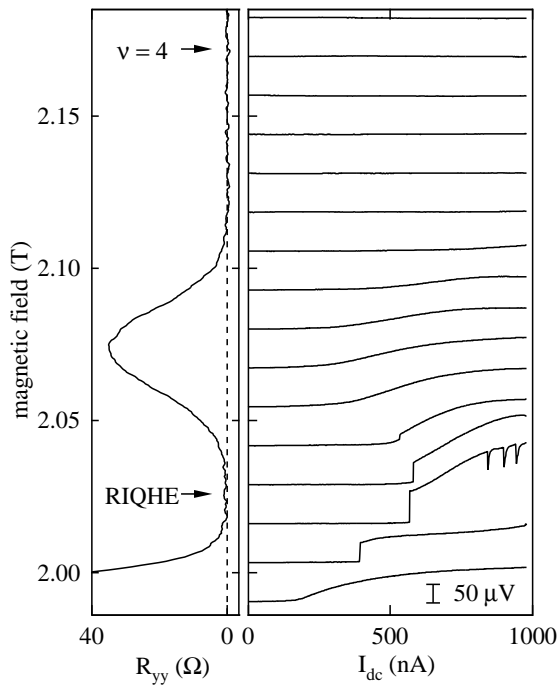}
\end{center}
\caption[figure 3]{Demonstration that I-V discontinuities are confined to the RIQHE.
Left panel: $R_{yy}$ vs. $B$ for sample B.
Right panel: Family of I-V curves, each vertically displaced to coincide 
with its magnetic field location on the resistance plot in the left panel.
}
\end{figure}

These sharp onsets of conduction are only seen in the immediate vicinity
of the RIQHE states. Figure 3 shows a sampling of I-V curves from sample
B (for clarity only sweeps with increasing current are shown). These data are
taken at equally spaced magnetic fields ($\Delta B = 13$mT) from the
low field side of the RIQHE near $\nu \approx 4 \frac{1}{4}$ to the 
center the $\nu=4$ integer QHE state. Also shown is the conventional resistance
$R_{yy}$ measured with a 10nA ac current excitation. It is clear from
the figure that the discontinuous jumps in the dc I-V
curves are only seen in the RIQHE state. Deep inside the ordinary
$\nu=4$ QHE state $V_{yy}$ remains zero, even out to much higher
currents than shown in the figure. Outside of the RIQHE, 
at most weak and smoothly varying non-linearities are found.
Essentially the same results are found in the vicinity of the $\nu
\approx 4 \frac{3}{4}$ RIQHE. Sample A shows non-linear behavior that is
quite similar to sample B, albeit with higher threshold currents.

\begin{figure}
\begin{center}
\epsfxsize=3.3in
\epsfclipon
\epsffile[104 260 511 598]{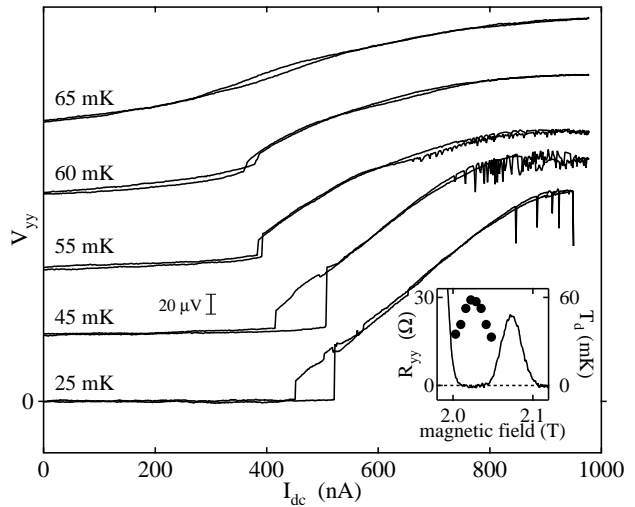}
\end{center}
\caption[figure 4]{I-V curves at several temperatures for sample B at $B = 2.02$~Tesla.
Curves are offset vertically for clarity. 
Inset: The step disappearance temperature $T_d$ (dots) and $R_{yy}$ at $T=25$mK
(solid line).}
\end{figure}

The qualitative difference between the I-V curves observed in 
the RIQHE and the conventional IQHE state is particularly
striking when the temperature dependence of the longitudinal resistance 
separating these regions is considered.  The small peaks in $R_{yy}$ 
and $R_{xx}$ shrink as the temperature is reduced. In sample B they
remain visible down to below T=25mK but in sample A, which has lower 
mobility, they essentially vanish. In spite of this, the discontinuities 
in the I-V curves are still found only in the narrow field range defined 
as the {\it re-entrant} QHE state via the resistance observed
at higher temperature.

The dc currents employed for Fig. 3 are large. Although the majority of
that current flows in the four edge channels of the $N=0$ and
1 Landau levels which lie below the Fermi level in the bulk of the
sample, the possibility of electron heating must be considered. Figure 4
shows that this is, perhaps surprisingly, not a serious problem. The I-V
curves in the vicinity of threshold are quite sensitive to temperature
down to below 25mK. As the temperature is raised, the conduction
threshold shifts to lower dc currents and disappears abruptly. In
the inset to Fig.~4, the temperature $T_d$ at which the
discontinuities disappear is plotted as a function of magnetic field. $T_d$
is largest ($\sim 60$mK) at the center of the RIQHE, and falls off
rapidly on either side.

Also apparent in Fig.~4 is the presence of noise in the conducting state
above threshold. Downward spikes in the voltage
are observed in the vicinity (magnetic field and
temperature) of where the discontinuity in the I-V is observed. When the
temperature and dc current are fixed, the spikes occur almost
periodically. The period, typically 1 to 40 seconds depending on 
temperature, decreases as $T$ approaches $T_d$. 
Above $T_d$ no voltage spikes are seen.  No spikes have been found, 
at any temperature, outside the RIQHE region.

These non-linear transport phenomena are suggestive of the
depinning and sliding transport of CDWs\cite{gruner}. In the present
case, the current flowing through the sample (via the filled Landau 
levels beneath the Fermi level) produces a Hall electric field 
transverse to the current.  This field exerts a force on the 
localized electrons in the uppermost Landau level. For small currents, 
this force is insufficient to delocalize these carriers and the longitudinal
resistance of the sample remains zero.  As the current increases these 
electrons eventually delocalize and the resistances become non-zero. 
If the electrons are individually localized by a random disorder potential, 
a gradual onset of conduction is expected. If, on the other hand, large 
collections of highly correlated electrons are pinned at a small number of 
sites, a much more abrupt transition to conduction is anticipated. Our data
reveal both kinds of behavior but only in the RIQHE are sharp onsets of
conduction observed. 

Similar ideas have been applied in the past to the 2DES at very high
magnetic fields where insulating behavior is observed\cite{shayegan}
in the lowest Landau level ($\nu \stackrel{<}{_\sim} 0.2$ in clean
electron systems). Non-linear transport
measurements\cite{goldman,williams,li,jiang} have indicated that
threshold fields (of varying sharpness) exist in this regime and many
have interpreted them as evidence for the depinning of an isotropic
Wigner crystal. In our case, where the insulating electrons are those in
a partially filled high Landau level, the spectrum of possible pinned
correlated states is broader and includes bubble\cite{fogler,moessner}
and stripe crystal\cite{fradkin} phases.

\begin{figure}
\begin{center}
\epsfxsize=3.3in
\epsfclipon
\epsffile[104 143 486 435]{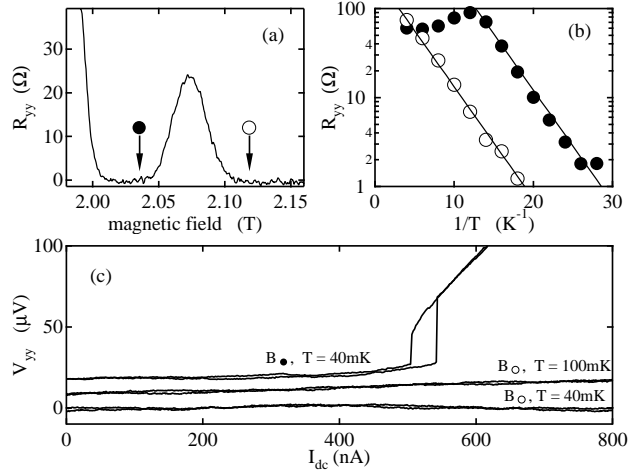}
\end{center}
\caption[figure 5]{Comparison of transport data just inside and just
outside the RIQHE state in sample B.  
Panel (a) indicates the magnetic fields where I-V
curves were taken.  In (b), an Arrhenius plot shows the temperature dependence
for the two fields, with lines to guide the eye.  In (c), I-V curves (offset
for clarity) show a discontinuous threshold only in the RIQHE state (top curve)
and not at the other field location (e.g. middle and bottom curves).}
\end{figure}

The non-linearity phenomena reported here are also
qualitatively similar to what has been seen in studies of the breakdown 
of the integer quantized Hall effect. Discontinuous I-V curves, hysteresis, 
and excess noise near the onset of conduction, have all been
reported\cite{nachtwei}. As Jiang, {\it et al.}\cite{jiang} emphasize in
their study of the insulating behavior of 2D electron systems at very
low filling factor, the observation of electric field thresholds in the
insulating state may be related to a thermal breakdown process, and not
the sliding of an underlying crystalline state. While thermal run-away
models\cite{nachtwei,komiyama} depend upon many factors, key among them
is simply the temperature dependence of the longitudinal resistivity. To
investigate the relevance of such models to the RIQHE, we compare the
temperature dependence of the resistance $R_{yy}$ at two filling
factors, one inside the RIQHE which shows a sharp conduction onset, and
one outside it which does not. These two filling factors, indicated in
Fig.~5a, are located symmetrically about the small peak in $R_{yy}$
which separates the RIQHE from the conventional $\nu=4$ QHE. At both
locations temperature dependence (Fig.~5b) of the resistance behaves in
an activated manner: $R_{yy}=R_0 exp(-E_A/T)$. While the measured
activation energies in the two cases are comparable ($E_A \sim 0.3$~K),
the prefactor $R_0$ is about 17 times larger at the point inside the
RIQHE than at the point outside it. At $T=40$mK, $R_{yy}$ is small
but non-zero at the RIQHE point and yet a sharp jump in the I-V curve was
still visible (Fig. 5c). In contrast, at no temperature were discontinuous 
(or even sharp) features observed at the other field location, a mere 
70mT away. This was true regardless of whether the actual $R_{yy}$ value 
was smaller, equal to, or larger 
than the value inside the RIQHE where a sharp conduction onset was observed. 
While these observations do not completely eliminate thermal run-away models
of the I-V discontinuities in the RIQHE, they do suggest that their origin 
lies elsewhere.

In conclusion, we report transport measurements of insulating states on
the flanks of the $N=2$ and higher LLs. In these regions the longitudinal
resistance vanishes and the Hall resistance becomes quantized at the nearby
IQHE value. The distinction of these RIQHE states from the conventional IQHE 
suggests that electron correlations are very important in their structure.
Measurements of current-voltage characteristics reveal discontinuous and 
hysteretic transitions between insulating and conducting phases of the electrons
in the uppermost LL. These dramatic non-linearities have only been found 
within the RIQHE and at very low temperatures ($T<60$mK). These findings are
highly suggestive of depinning of charge density waves, but other mechanisms 
may yet prove responsible.

This work is supported by the National Science Foundation and the Department of
Energy.

\end{document}